%% file: main.tex
\documentclass[sigconf]{acmart}

\usepackage{algorithm}
\usepackage{algorithmic}
\usepackage{subfigure}
\usepackage{comment}
\usepackage{url}

\AtBeginDocument{%
  }

\setcopyright{acmcopyright}
\copyrightyear{2018}
\acmYear{2018}
\acmDOI{XXXXXXX.XXXXXXX}

\begin{document}

\title{Cluster Resource Management for Dynamic Workloads\\ by Online Optimization}

\author{Nader Alfares}
\affiliation{
    \institution{Pennsylvania State University}
    \city{University Park, PA}
    \country{USA}}
\email{nna5040@psu.edu}

\author{George Kesidis}
\affiliation{
    \institution{Pennsylvania State University} 
    \city{University Park, PA}
    \country{USA}}
\email{gik2@psu.edu}

\author{Ata Fatahi Baarzi}
\affiliation{
    \institution{LinkedIn} 
\country{USA}
}
\email{immrata@gmail.com}

\author{Aman Jain}
\affiliation{
    \institution{Microsoft} 
\country{USA}
    }
\email{aman.jain@microsoft.com}


\begin{abstract}
Over the past ten years, many different approaches have been proposed for different aspects
of the problem of resources management for long running,
dynamic and diverse workloads such as processing query streams 
or distributed deep learning. Particularly
for applications consisting of containerized microservices, researchers
have attempted to address problems of dynamic selection of, for example:
types and  quantities of virtualized services (e.g., IaaS/VMs), 
vertical and horizontal scaling of different microservices, 
assigning microservices to VMs, task scheduling,
or some combination thereof. In this context,
we argue that frameworks like simulated annealing 
are highly suitable for online navigation of trade-offs between
performance (SLO) and cost, particularly when 
the complex workloads and cloud-service offerings vary over time.
Based on a macroscopic objective that combines both performance and cost terms, annealing facilitates light-weight and coherent policies of exploration and exploitation.
In this paper, we first give some background on simulated annealing and then experimentally
demonstrate its usefulness for different case studies,
including service selection for both a single type of workload (e.g., distributed deep learning) and a mixture of workload types (exploring a partially categorical set of options),
and container sizing for microservice benchmarks.
We conclude with a discussion of how the basic annealing platform
can be applied to other resource-management problems, hybridized with
other methods, and accommodate  user-specified rules of thumb.
\end{abstract}

\settopmatter{printacmref=false}
\setcopyright{none}
\renewcommand\footnotetextcopyrightpermission[1]{}
\pagestyle{plain}

\maketitle

\input{intro}
\input{background}

\section{IaaS Procurement Approach}\label{sec:annealing}

Consider a long job stream whose composition and workload profile may change periodically over time. This change could be due to changes in the types of jobs and their proportions and/or the datasets on which the jobs operate. A goal here is to decide on the most performance and cost effective IaaS cluster, including consideration of autoscaling costs, dynamic changes
to the effective capacities and prices of services considered, or new service offerings.

First, consider a stream of the same type of job, e.g.,
the jobs are described by the same DAG of component tasks, for
a fixed cluster.
Let $x_{k}$ be the cluster configuration for the $n^{\rm th}$ job (e.g., indicating the total number of cores in the cluster). Starting with a random configuration for $x_0$, we apply annealing over the course of the job stream with minimizing objective
$$Y_n = t_n + \lambda c_n$$  
where $t_n$ and $c_n$ respectively are the total execution time and a ``cost" for job $n$ under the currently chosen service configuration. The user specified term $\lambda>0$ weighs the cost against the execution time. 

Second, we formulate the objective to be able to consider a blend of multiple types of jobs. The composition of the blended workload can be described using weighted averages of each type. For example, consider a workload that consists of  $N$ types of jobs. The minimizing objective for this workload can be described as
$ Y = \sum_{i=1}^{N} \alpha^{(i)}  t^{(i)} + \lambda C,$
where $\alpha^{(i)}>0$ is the weight for workload type $i$ ($\sum_{i=1}^{N} \alpha^{(i)} = 1$),  $t^{(i)}$ is a measure of its performance to be minimized, and $\lambda>0$ is a parameter relatively weighing
a measure of the cost of the cluster, $C$.
The user-specified parameter $\alpha^{(i)}$, which can be interpreted as the relative
priority of workload type $i$, may change dynamically as the workloads experience variations over time. 
Upon arrival of a new job (or new set of jobs) $n$, 
we run it (them) with the configuration $z_n$
$$z_n = x_{n-1} + e_v$$
where $e_v$ represents a possible
``step size" (incremental change) when exploring configurations
and $x_{n-1}$ represents the current ``accepted" configuration.
We set
$x_n=z_n$ (i.e., accept the configuration $z_n$) with probability
$$ \exp(-\max\{Y_n-Y_{n-1},0\}/\tau)$$
where
$Y_n$ is the objective evaluated for job $n$, and
$\tau$ is the temperature parameter controlling the 
degree of exploration and exploitation;
otherwise $x_n=x_{n-1}$.

\begin{figure}[b]
    \centering
    \includegraphics[width=0.45\textwidth]{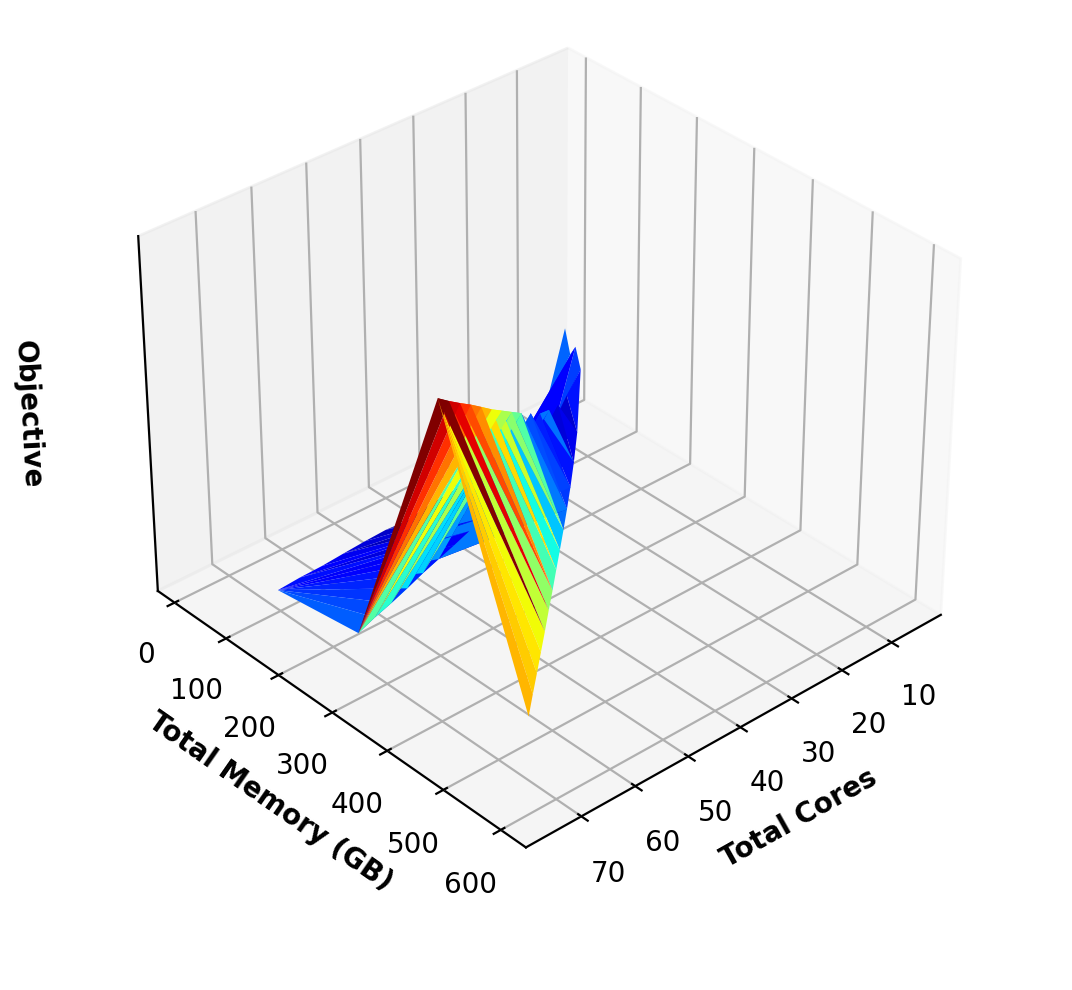}
    \caption{Blended workload characteristic based on four types of EC2 instances: General Purpose, Compute Optimized, Storage Optimized, and Memory Optimized.}
    \label{fig:blend_ec2}
\end{figure}

\begin{figure}
    \centering
    \includegraphics[width=0.4\textwidth]{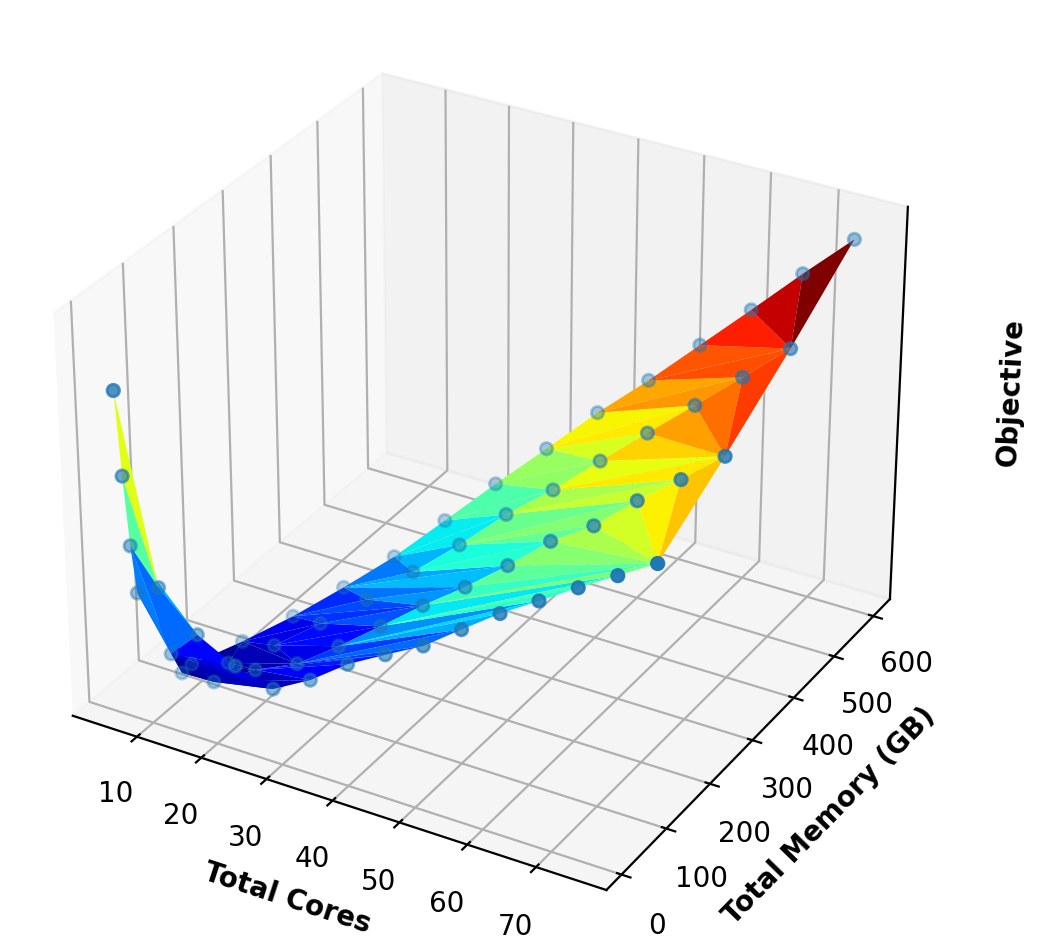}
    \caption{Blend of HiBench workload types (Wordcount, K-means and PageRank) with adjustment for storage-optimized instances for better comparisons.}
    \label{fig:eval_blend}
\end{figure}

\begin{figure*}
    \subfigure[Low Temperature Annealing ($\tau = 10$)]{
        \includegraphics[width=0.3\textwidth]{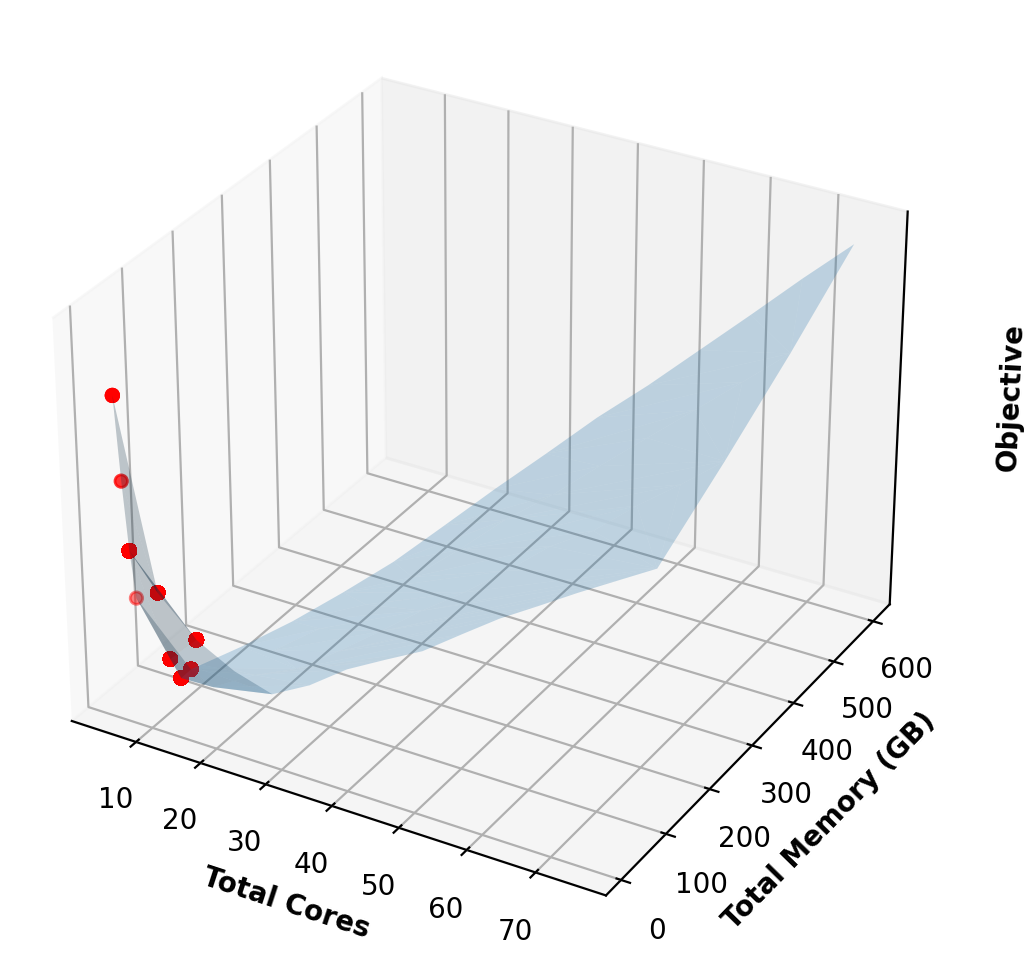}
        \label{fig:eval_low_temp}
    }
    \subfigure[Medium Temperature Annealing ($\tau = 25$)]{
        \includegraphics[width=0.3\textwidth]{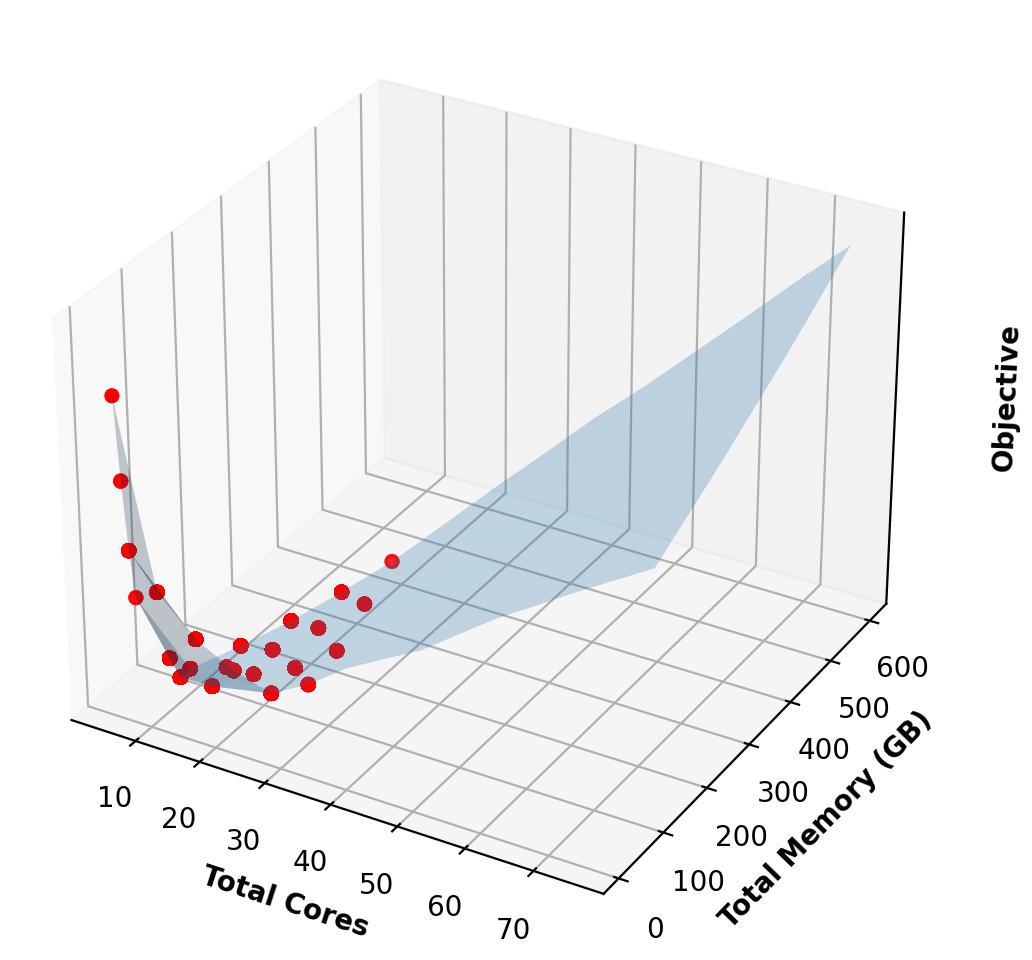}
        \label{fig:eval_med_temp}
    }
    \subfigure[High Temperature Annealing ($\tau = 50$)]{
        \includegraphics[width=0.3\textwidth]{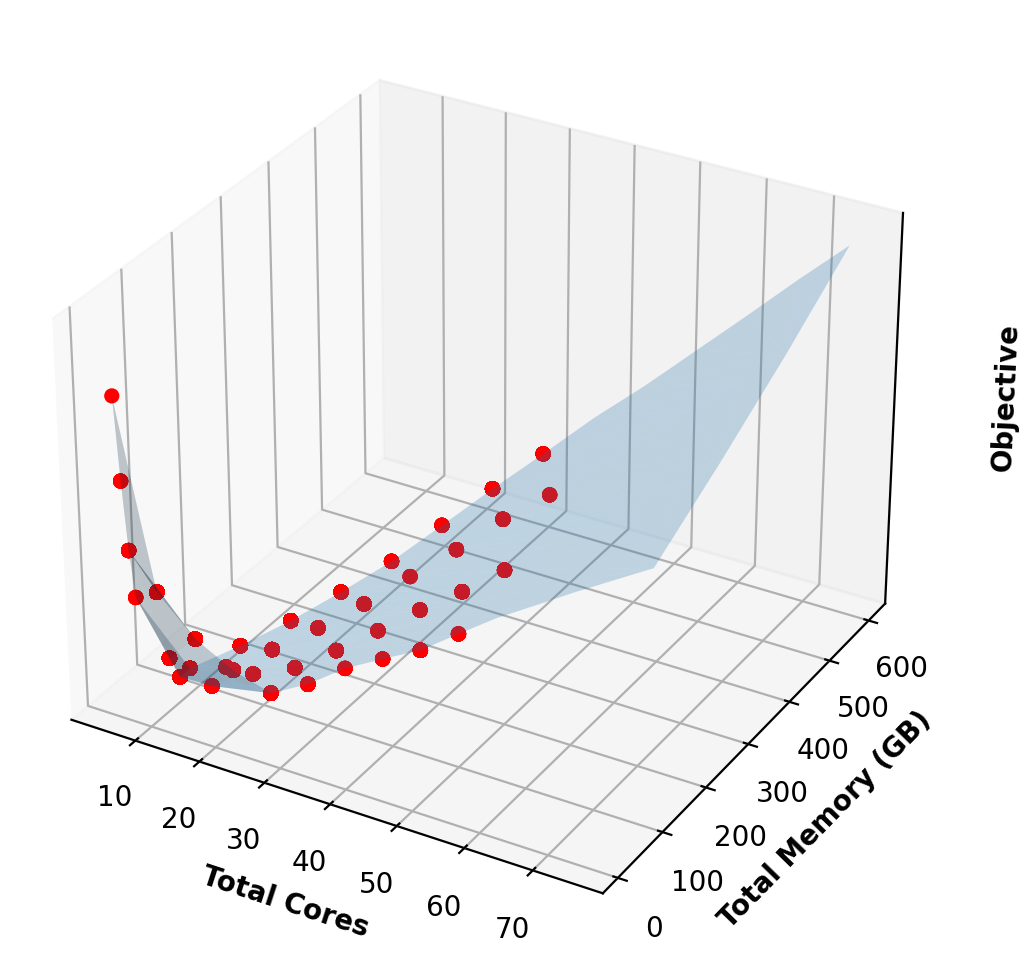}
        \label{fig:eval_high_temp}
    }
    \caption{Performing annealing on a blended workload of three types of jobs: Wordcount, K-means, and Pagerank}
    \label{fig:eval_temp}
\end{figure*}

\section{Experimental Evaluations - IaaS Procurement}\label{sec:expts}
\subsection{Experimental Set-Up}
\label{sec:expts_setup}
We evaluate our simulated annealing approach using CloudLab \cite{cloudlab} machines. Our set-up consists of six rs630 nodes (a master node and 5 worker nodes). Using Apache Spark (3.1.2v), we exploit the Spark configuration file in the master node to set the new desired configuration when performing annealing.

\subsection{Modeling Cost of EC2}
\label{sec:EC2_Modeling}
We reference AWS's EC2 per core pricing for different on-demand instance types. For each instance type, a 
pre-defined memory allocated per core is assumed (e.g., m6g.medium instances allocate 4 GB of memory per core procured). We also consider hypothetical instances ``between'' those offered by AWS with corresponding price adjustments (further details are given in section \ref{sec:single_jobs}).

\subsubsection{Each job executed independently}
\label{sec:single_jobs}
Using a blend of three different type of HiBench jobs \cite{Huang2012HiBenchA} (we set the weight factor $\lambda = 5$, which we empirically found suitable for our experiments), we evaluate the objective values, under different configurations, for the blended workload as described in section \ref{sec:annealing}. We generate a stream of independent jobs and record their average execution times under configurations provided by the annealing process, i.e., number of VM instances of a certain type where each VM  type is characterized by a number of cores and memory allocated per core. Figure \ref{fig:blend_ec2} depicts the objective values for different configurations of such blended workload. The peaks in figure \ref{fig:blend_ec2} evaluate the objective values when using storage-optimized instances\footnote{Latency to local storage was not a significant performance factor in our experiments. Hence, AWS instances that use Elastic Block Store (EBS) are also emulated using SSDs, but with their actual AWS pricing.}. 
Note that different ordering of the {\em categorical} instance types in defining the search space may introduce local minimums that are not global. Annealing can overcome local minima more quickly at higher temperatures $\tau$.



\begin{figure}
    \centering
    \includegraphics[width=0.4\textwidth]{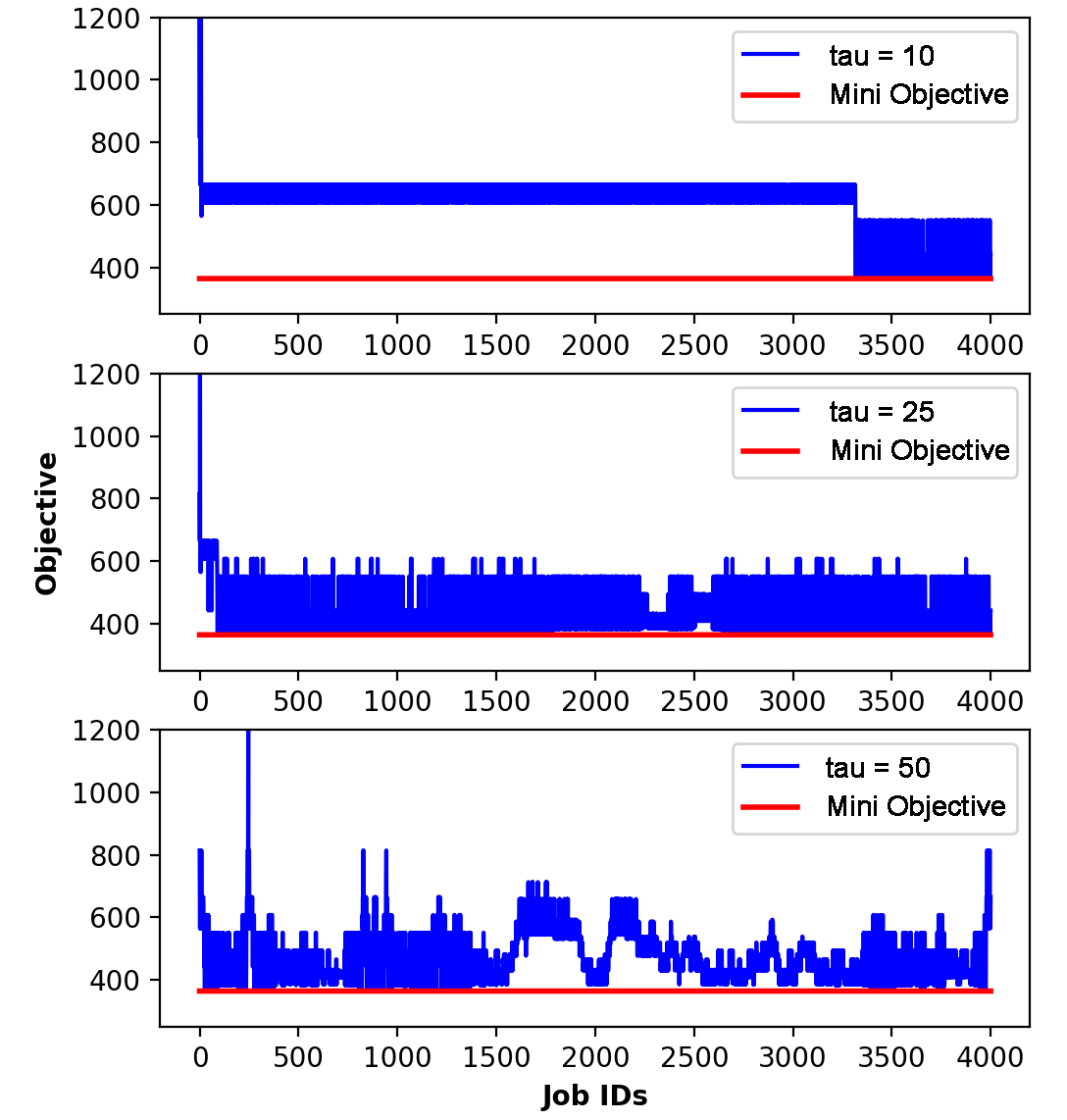}
    \caption{The occurrences of exploitation and exploration depends on the temperature. Each chart in the above represent performing an annealing process under a fixed temperature for a blended workload.}
    \label{fig:eval_temp_2}
\end{figure}

\begin{figure}
    \centering
    \includegraphics[width=0.75\columnwidth]{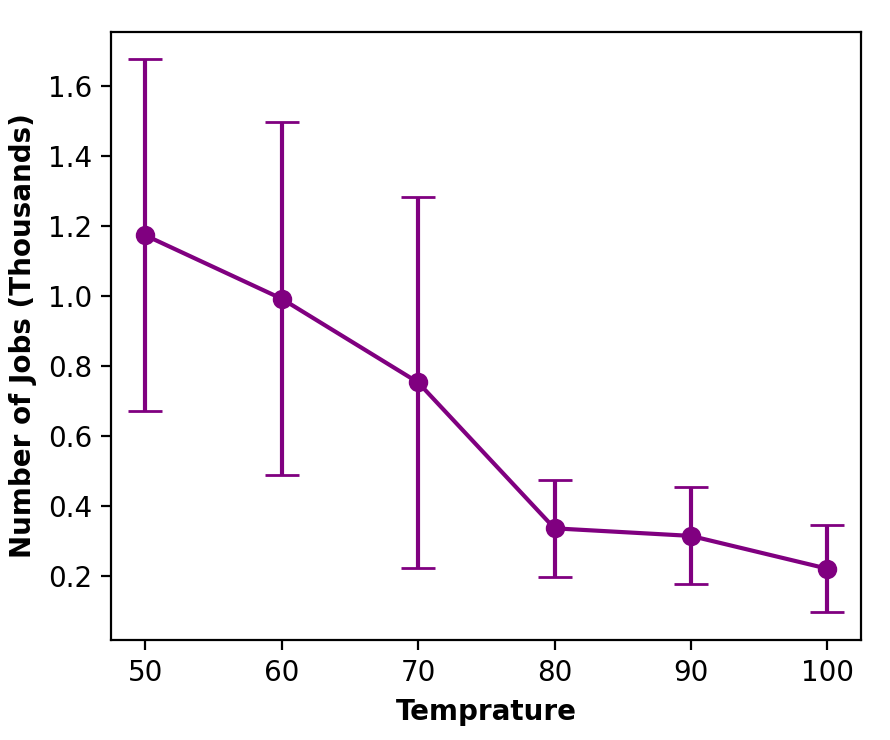}
    \caption{The number of jobs submitted before a configuration with a minimum objective is selected depends on the temperature. The vertical
    bars represent 95\% confidence intervals ($\pm 2$ sample standard
    deviations).}
    \label{fig:eval_find_min}
\end{figure}

We replace the pricing of storage-optimized instances with a hypothetical family of instances for better comparisons (i.e., all families of instances have similar local storage performance) shown in figure \ref{fig:eval_blend}.
Figure \ref{fig:eval_temp} shows different simulations of jobs streams while performing annealing under different temperatures. The red dots represent  different service-cluster configurations. From figures \ref{fig:eval_temp} and \ref{fig:eval_temp_2}, we can observe that more exploration is performed by annealing as the temperature $\tau$ increases. 
Furthermore, figure \ref{fig:eval_find_min} shows that as the temperature increases,   the annealing process more quickly finds the configuration that minimizes the objective\footnote{In this paper, in order to assess the annealing mechanism, we separately identify by exhaustive search the configurations that globally minimize the objective under consideration.}, but with greater service variation due to higher service exploration.


\subsubsection{Jobs executed in parallel}

The annealing approach can also be employed for workloads that consist of jobs that are executed in parallel (i.e., when jobs compete for resources) and a job queue may be present. The minimizing objective can be adjusted for this case by measuring the total sojourn time of jobs instead of just the execution times. The annealing process performs similarly to what is described in section \ref{sec:single_jobs}.

\subsection{Adaptation of Annealing}
\begin{figure}
    \centering
    \includegraphics[width=0.45\textwidth]{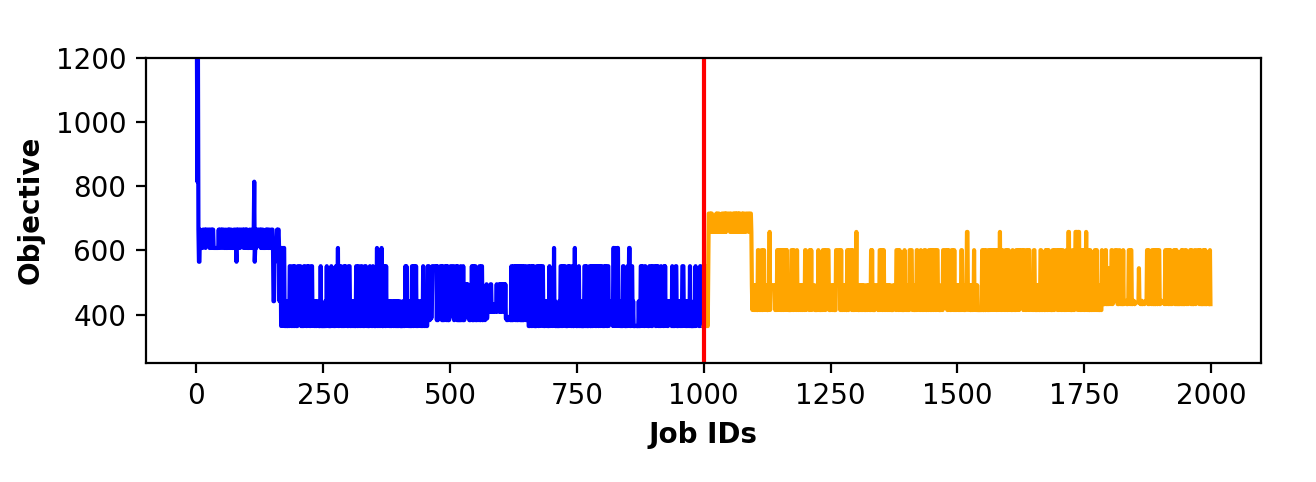}
    \caption{Annealing adapting to change in the blended Hibench workload.}
    \label{fig:HiBench_adaptation}
\end{figure}

We evaluate the adaptation of annealing for dynamic changes in the blended workload. Figure \ref{fig:HiBench_adaptation} shows the computed objective values for a stream of blended HiBench jobs. Here, we portray the change in workload as a change in the distribution of the blend (red vertical line depicts the point of time when the change occurs). The blue lines depict the objective values computed before the change occurs, while the orange lines depict after such. The oscillation in the computed objective values is due to the exploration nature of annealing at a positive temperature.
After the change, we observe that annealing adapts to the change by finding the new 
objective-minimizing service configuration through exploration as well.

\subsection{Deep Learning Workload}
\begin{figure}
    \centering
    \includegraphics[width=0.45\textwidth]{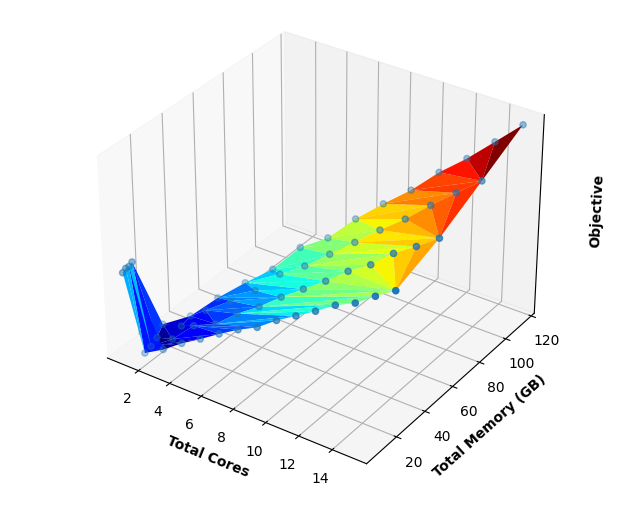}
    \caption{Training a Deep Neural Network using MNIST dataset for handwritten digits recognition.}
    \label{fig:MNIST}
\end{figure}

\begin{figure}
    \centering
    \includegraphics[width=0.45\textwidth]{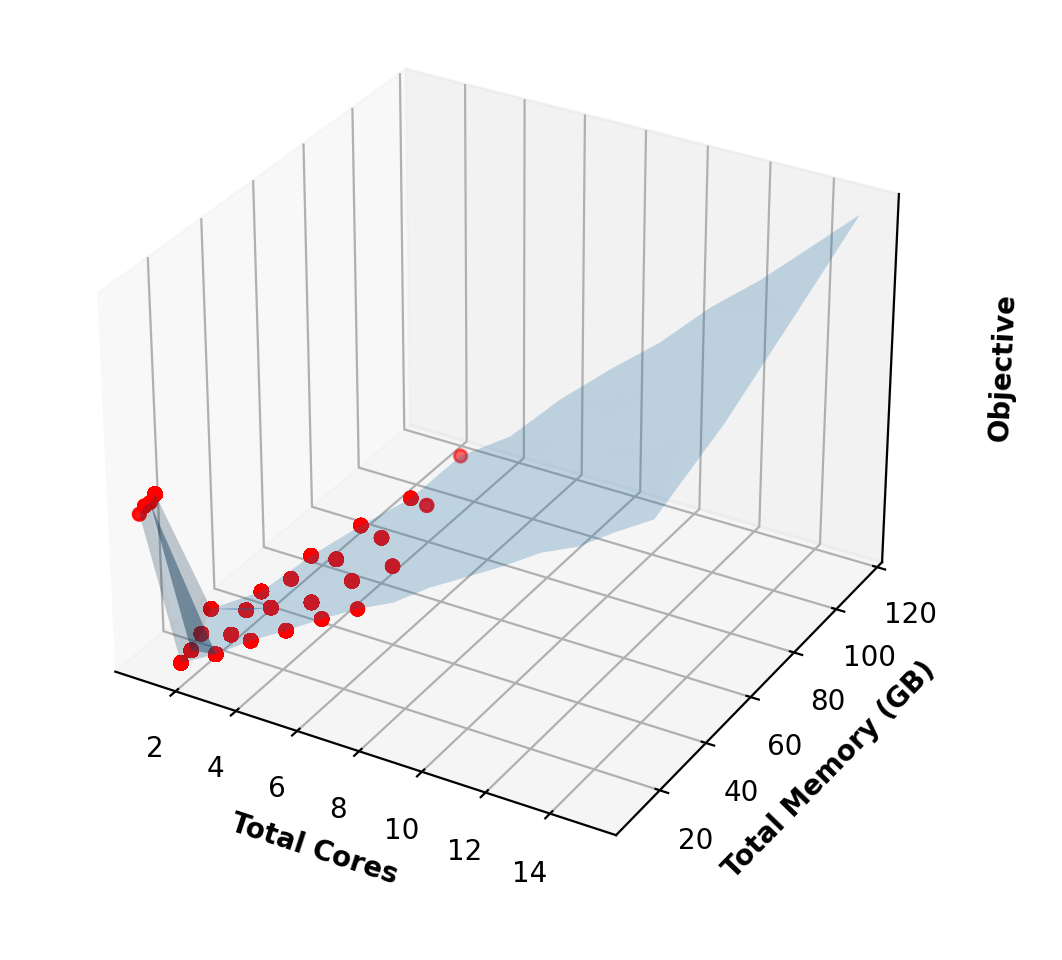}
    \caption{Performing annealing on a Deep Neural Network using the MNIST dataset.}
    \label{fig:MNIST_temp_1}
\end{figure}

\begin{figure}
    \centering
    \includegraphics[width=\columnwidth]{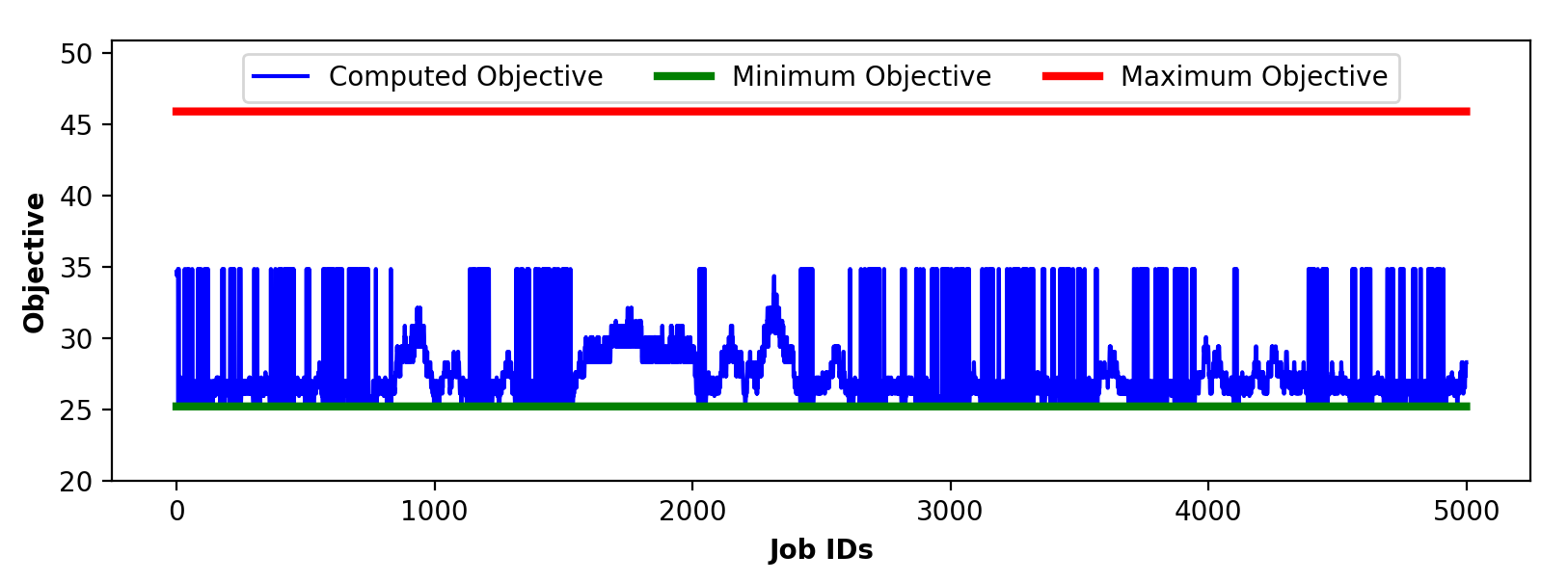}
    \caption{Computed objectives for selected configurations by the annealing process (blue line). The red and green lines correspond to configurations with the maximum and minimum objective values, respectively.}
    \label{fig:MNIST_objs}
\end{figure}

The annealing method can also be utilized in training Deep Neural Networks (DNNs), i.e., deep learning. Figure \ref{fig:MNIST} depicts the characterization of distributed deep learning,
on our Spark cluster,
using the Keras library \cite{keras} to train a Convolutional Neural Network (CNN)
to recognize handwritten digits of the MNIST dataset \cite{MNIST}. 
Using the pricing model described in section \ref{sec:EC2_Modeling}, the goal is to find the service configuration that minimizes the objective function as described in section \ref{sec:annealing} (we set the weight factor $\lambda = 1$). The execution time here refers to training a model for one epoch. In our experiments, we set the initial configuration to 1 core with 4 GB of memory. Figure \ref{fig:MNIST_temp_1} shows the selected configurations that were explored by the annealing process (shown as red dots). From figure \ref{fig:MNIST_objs}, we find that our annealing approach is capable of quickly finding the configuration that minimizes the objective.

\subsection{Comparison with a Deep Learning Approach}

Supervised training of a DNN (a.k.a. deep learning) heavily relies on collection and curation of a large corpus of labeled data. This process typically involves acquiring a representative dataset that captures a significant exploration of the problem domain's characteristics and associated ground truth labels. 
The overall cost of the data acquisition process is rarely reported (though the computational
cost associated with the deep learning process itself is commonly reported).
Moreover, there typically are a large number of hyperparameters associated with the DNN architecture and the deep learning process itself (though it's often the case that defaults are chosen without consideration of the application domain). 

On the other hand, simulated Annealing (and other light-weight online optimization methods including hybrids) typically does not involve offline data collection. Instead, it only online explores the solution space by iteratively adjusting the solution based on a given objective function and the acceptance of new solutions.

An important issue regarding supervised learning in complex and time-varying environments is that
there may be gaps in its training resulting in poor decisions made when the operating conditions 
change to an unexpected state, i.e., ``model drift." Detection of model drift may trigger a ``mini" exploration phase and subsequent adaptation/refinement of the supervised learning model.
In an effort to compare two such approaches under model drift
without delving into the hyperparameters of deep learning, 
we consider online random search to explore enough data samples before finding an acceptable solution. We compare the data collection phase for deep learning to the exploration phase in simulated annealing after a significant change in operated state.   In summary, by modeling the data collection phase as random search, akin to simulated annealing's exploration phase, we can establish a simple basis for comparison and evaluation between simulated annealing and generic supervised learning in response to model drift.

\begin{figure}[h]
    \centering
    \includegraphics[width=0.45\textwidth]{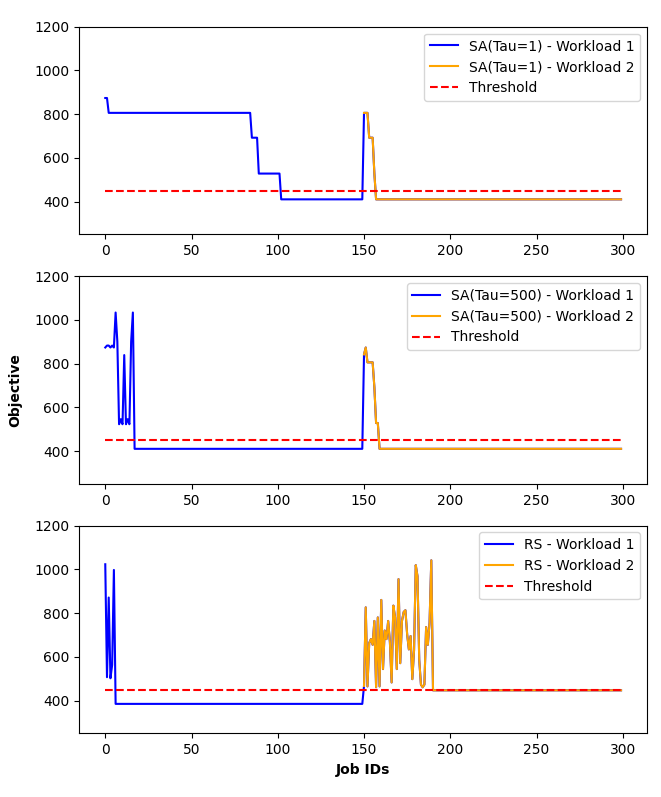}
    \caption{Deep Learning approach comparison: SA with low temperature 1 (top), SA with high temperature 500 (middle), and RA (bottom). The initial workload had 
    70\% Wordcount, 20\% K-means and 10\% PageRank. The workload at iteration 150 changes to
      20\% Wordcount, 70\% K-means and 10\% PageRank.}
    \label{fig:SA_DL}
\end{figure}

Figure \ref{fig:SA_DL} represents the adaptation of Simulated Annealing (SA) and Random Search (RS) to a change in the workload mixture for the problem of VM service selection. In addition, both approaches adhere to a candidate solution without exploration when the cost objective drops below a (user specified) threshold.
Again, here the initial RS adapts only to the current workload only so it's actually an online approach like SA -- forming a training dataset for a DNN would require much more initial exploration for a variety of different workloads not just the current one as depicted in the figure at iteration zero.

Note from Figure \ref{fig:SA_DL}, high-temperature SA finds an adequate solution at about the same time as RA for the initial workload during the iteration interval $[0,149]$.
When the workload changes at 150 causing the cost objective to rise above the threshold,
RA takes longer to find an adequate solution compared to both high and low temperature RA because the adequate solution found by SA was similar to that of the initial workload.

\subsection{Discussion: Adaptive Temperature Adjustment in Simulated Annealing}

When a running average of the cost objective value improves, indicating progress towards the desired optimization goals, the temperature can be reduced. This reduction in temperature corresponds to a decrease in the likelihood of accepting worse solutions, leading to a more focused (depth of) search around the currently nearby optimum. 
Several different strategies can be employed for adaptive temperature adjustment in SA, depending on the nature of the optimization problem and the specific cost and performance metrics involved, for example:
\begin{itemize}
\item Continuous Logarithmic: The classical logarithmic (slow) cooling schedule \cite{AK89}.
\item Threshold Based: Predefined thresholds for the objective value are established with temperature changes at each threshold. If the current objective value crosses a threshold, the temperature is reduced to encourage exploitation. Alternatively, if the cost objective exceeds a certain threshold, the temperature may be increased to encourage exploration. The method of the previous subsection is an example.
For another example, at any given threshold, temperatures can increase exponentially  (e.g., $\tau\rightarrow 2\tau$) 
and decrease additively  
(e.g., $\tau \rightarrow (\tau-1)^+$).
\item Iteration Based: The temperature adjustment is based on the number of iterations or the convergence behavior. For instance, if no significant in the cost objective is observed after a certain number of iterations, the temperature can be incremented to encourage further exploration. 
\end{itemize}
For all of the above, annealing may recall the {\em most recent} adequate solution $x$, and may return to $x$ when temperatures increase, bearing in mind that $x$ may not be adequate  for the current operating conditions.
Also, some or all of the decisions of the above methods could also be based on other statistics such as running average mean (rather than the current value) or variance  of the cost objective.

\section{Experimental Evaluation - Container Sizing for Microservices}\label{sec:expts-mus}

\begin{figure*}[ht]
    \centering
    \includegraphics[width=\textwidth]{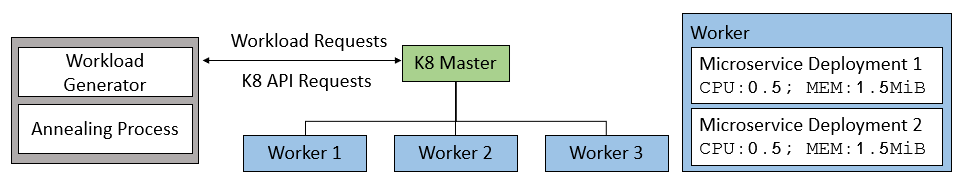}
    \caption{Container sizing experimental setup.  }
    \label{fig:MS_Diagram}
\end{figure*}

\subsection{Experimental Set-Up}
For our container sizing experiments (Figure \ref{fig:MS_Diagram}), we use CloudLab \cite{cloudlab} to set up a cluster of r320 nodes. Specifically, we configure the cluster to have five nodes, with one node serving as the Kubernetes master and three nodes as Kubernetes workers. Additionally, we designated one node as the workload generator, which also includes the annealing process that communicated with the Kubernetes API at the master node to adjust and modify the resources allocated (configurations) to the deployments. We run our experiments on two microservice applications provided by DeathStarBench \cite{deathstarbench} (using helm-chart): \textit{Social Network} and \textit{Hotel Reservation}. The former is an implementation of a social media network that allows users to read, post, and react to social media posts, whereas the latter is an implementation of a browser-based application that allows users to search and make reservations for hotels. We employ the annealing approach to size the resources allocated to microservices' deployments. We generate the workload for the two microservices using \texttt{wrk2} \cite{wrk2}, along with workload generator scripts (the benchmark provides both).


\subsection{Defining the Annealing Objective}
Unlike service selection (section \ref{sec:expts}), here we define the objective function as
$Y_n = t_n + \lambda (U_{\rm cpu} + U_{\rm mem})$
where $t_n$ is the average execution time for epoch $n$, $U_{\rm cpu}$ and $U_{\rm mem}$ are CPU and memory utilization, respectively, and again $\lambda > 0$ is a user-specified factor that weighs the average latencies of the requests against the utilization of resources by the microservices. For experiments of this section, we set $\lambda = 1$, which we empirically found to be suitable for our demonstrations.

\subsection{Experimental Results}
For each experiment, we deploy the microservice application to the Kuberentes cluster while setting the resource request and limit to the minimum configuration for cores and memory for each deployment (we set the minimum for CPU and memory to be at 0.1 CPU and 0.1 MiB). Second, we set up our workload generator to send requests at one-minute intervals (epochs) and measure the mean latency of the requests. After every epoch, the annealing process evaluates the objective value and explores a random microservice to patch with a new resource request and limit at steps of 0.1 CPU/MiB for CPU/memory. When maximum resources are allocated, the annealing exploration phase can either 
modify
resources allocated to a random microservice, or reallocate resources from one microservice to another.

\begin{figure}[ht]
    \centering
    \includegraphics[width=\columnwidth]{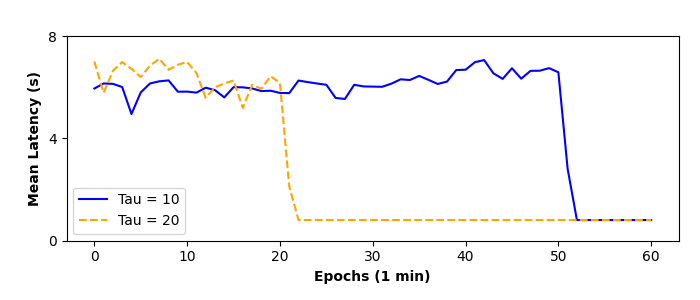}
    \caption{Social Network: a configuration with a minimizing objective is found quickly at a relatively high temperature.}
    \label{fig:DS_SM}
\end{figure}

\begin{figure}[ht]
    \centering
    \includegraphics[width=\columnwidth]{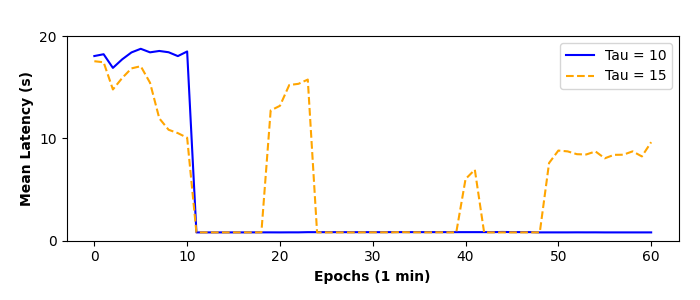}
    \caption{Hotel Reservation: an optimal configuration is retained at a longer period of time for annealing with relatively low temperatures.}
    \label{fig:DS_HOTEL}
\end{figure}

Figure \ref{fig:DS_SM} depicts the response times of the read requests to the social network application while performing annealing at different temperatures. The figure shows that a minimizing configuration of the microservices resources can be quickly discovered at a relatively high temperature.  In practice, when a configuration with an acceptably small evaluated objective is found, the temperature can be lowered, retaining the configuration desired for a longer period of time, as depicted in figure \ref{fig:DS_HOTEL}.

\begin{figure}[ht]
    \centering
    \includegraphics[width=\columnwidth]{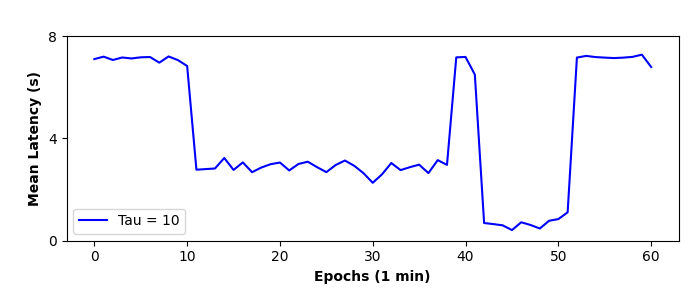}
    \caption{Social Network: performing annealing }
    \label{fig:DS_SM_2mins}
\end{figure}

Figure \ref{fig:DS_SM_2mins} depicts the response time of read requests made to the social network application while performing annealing at temperature $\tau = 10$. This experiment describes a scenario where an annealing process selects and remains at a configuration that minimizes the objective value (e.g., a local minimum) for a period of time before finding the optimal configuration through exploration.  Note that to find the globally minimizing configuration,  configurations that {\em increase} the objective are first explored.

\section{Discussion}



This paper describes a simulated-annealing approach to the management of the performance and cost (resources) of a cluster of virtual machines in the public cloud which is agnostic to both services and dynamic workload. 
Online annealing has very low complexity and, at moderate temperatures, is responsive to changes in the workload or available service suite (in contrast to approaches based on deep neural networks which require a very large training dataset and costly deep learning and offline reinforcement learning processes). 
Considering its ease and low cost of implementation and reasonably good performance, 
simulated annealing and its variants form a compelling class of baseline 
frameworks for online resources management of clusters.

In this paper, 
to demonstrate its ease of deployment and use and its good performance,
we evaluated an implementation of annealing for
\begin{itemize}
    \item service selection, evaluated using Apache Spark implementations of HiBench and distributed deep learning workloads, and costs based on AWS VM instance pricing, and 
    \item pod or container sizing, i.e., 
    vertical scaling, evaluated using Kuberenetes (K8s) for microservice workloads, with costs based on 
    efficiency of use of the existing instances.
\end{itemize}
Note that service selection can be expanded to consider the
possible benefits of serverless functions, 
e.g., \cite{splitserve19}, 
particularly for stateless microservices,
or to consider different storage options particularly for data intensive tasks.

As mentioned in Section \ref{sec:intro},
online simulated annealing can also be (jointly or
independently) used:
\begin{itemize}
    \item to adjust the task-scheduling policy;
    \item for assigning containers or pods to VMs (rather than K8s' default ``bin packing"
    \cite{kube-bin-packer});  or
    \item for horizontal scaling of microservices (i.e., creation of service replicas),
    \end{itemize}
e.g.,\cite{Gan2010, Pandit2014, Liu2016, Telenyk2017,Attiya17, Laarhoven1992}.
All such annealing mechanisms can work with a common temperature parameter or separately adjustable ones, and some obviously will need to operate at different time-scales: fast (task scheduling), moderate (container sizing, assigning containers to VMs), slow (horizontal scaling), slowest (service selection).

Rules of thumb can be easily incorporated into 
simulated annealing neighborhood function $\nu$, e.g., to constrain assignment of service replicas to the same VM to take advantage of statistical multiplexing \cite{wisefuse}, 
or to require that certain microservices be assigned to VMs with GPU support,
or to constrain assignment of certain groups of different microservices to the same VM to take advantage of data locality, 
where the last type of requirement may also constrain horizontal scaling. 

The foregoing discussion is \textbf{not} meant to suggest that annealing, or some other method of automatic 
online optimization, needs to be applied to task scheduling or for the assignment of containers/pods to VMs.
The default ones provided by, e.g., Spark and K8s (and used in our experiments) may be adequate in some use cases.
Note that under K8s, such policies are user configurable
\cite{kube-scheduler,kube-priority,kube-scheduler-extensions,kube-bin-packer,kube-affinity}. That is, the K8s API can
be used to create affinities between certain
types of VMs and certain microsesrvices and
among microservices. Moreover, the K8s API can be used to configure the ``bin packer" so that, e.g., the fewest possible VMs are used 
(allowing some VMs to be released to save cost),
or to balance the pods among the existing VMs, subject to any stipulated constraints.

Simulated annealing  (or another online optimization framework like 
Bayesian optimization \cite{Garnett23}) 
can easily be modified, e.g., to remember the best states visited in the recent past and return to them with lower temperature if ``exploration" at higher temperatures has not recently yielded good results (smaller objective). Also, simulated annealing can be hybridized with other known methods, e.g., with random restart or genetic algorithms to improve breadth of search (exploration), or by excluding recently visited states to improve efficiency of depth of search (exploitation).

The scripts we developed for our Spark and K8s based experiments are posted on GitHub 
\cite{OSA}.

\section*{Acknowledgements}
This  work was supported by two NSF grants 2212201 and 2122155.

\bibliographystyle{plain}

\def\cprime{$'$}

\end{document}

%% file: intro.tex
\section{Introduction and Motivation}\label{sec:intro}

Resource management in the public cloud typically involves selecting available services spanning compute, storage and networking to minimize cost (or ``cloud spend" \cite{rightscale2017survey}) subject to workload  
Service-Level Objectives (SLOs, i.e., performance requirements).
The problem is particularly challenging when serving a plurality of complex time-varying workloads. 
For a given set of job streams,
the optimal service suite may involve a cluster composed of a variety of services.
Even within services of the same type, resource instances can vary greatly. For example, VMs can have different amounts and types of memory, CPU cores,  hardware accelerators, cross connects (e.g., PCIe, NVLink), and networking. 
In some cases, users can optionally pay to collocate VMs on the same physical server or to provision networking resources connecting their VMs to storage.\footnote{There are similar efficient resource allocation challenges in  ``bare metal" and private data center scenarios where virtualized public-cloud services are not in play.}

Deploying application software using a
microservice architecture, e.g., \cite{deathstarbench,firm20,SARA20,Fahmy21},
has the advantage of modular code design and maintenance.
Some microservices can be used by different applications. Also,
each microservice can be executed in a separate, light-weight
container. 

Particularly in an edge cloud setting where operating costs are high,
the user may wish to economize by explicitly
provisioning containerized microservices (vertical scaling)\footnote{In \cite{firm20}, it's argued that provisioning containers can prevent resource starvation of bottleneck microservices toward improving total job execution times, compared to simply relying on the VM's OS to schedule component tasks as they arrive in a ``best effort" fashion without considering end-to-end job execution times.}
and replicating  microservices (horizontal scaling) as necessary.
Data locality issues may be involved in how containers are assigned
to VMs. 
Different task scheduling policies have been proposed 
which can be implemented on cluster managers such as Kubernetes (K8s) \cite{kube-scheduler,kube-bin-packer}.
In addition, a default task scheduler and a task assignment policy to containerize microservices can be augmented with simple rules of thumb, e.g., to exploit data locality  \cite{wisefuse} and to address persistent straggler tasks 
\cite{Mohan21}\footnote{Also, rather than relying on a default mechanism 
of a cluster manager (e.g., K8s), a user may wish to control how
microservices are ``packed" into the VMs of their cluster, to both
economize on the number of VMs and, again, to address data locality issues
as a job's tasks are executed as a sequence of microservices.}.

For complex, diverse and dynamic workloads incident to a cluster,
deciding a cloud service suite and a container-sizing and replication policy 
is a difficult task (even if, e.g., a default task scheduling is used), and a typical user (cloud tenant/customer) may provide little more guidance than ``macroscopic'' SLOs and a cost budget. 
Some approaches to this problem 
involve using a deep (large) artificial neural network (DNN) 
to determine a service suite, size containers, schedule tasks, etc.,
e.g., \cite{Delimitrou16,Alizadeh18,firm20}.  However, deep learning is supervised requiring a very large curated training dataset representing a huge group of ``exploratory" experiments for a given workload \cite{Higginbotham19}.
Unanticipated
changes to the workload (even to just the dataset on which it operates) or to the offered service suite (i.e., ``model drift'') 
may require identifying and labelling
new operational ``samples" to periodically refine the DNN, which itself may require
significant computational cost.

Recently, researchers and practitioners have explored the use of 
 other traditional but more light-weight and adaptable
 means of decision-making (including using model-based adaptive control, PID control, and Markov decision processes) where DNNs play only a partial role at most, e.g., \cite{google-autopilot,Caerus21}.
 For example, particle-swarm type optimization has been used for
 load balancing, e.g., \cite{Singhal20}.
Also, genetic
algorithms (GAs), e.g., \cite{WW94,wilcox2011solving}, have been used
to explore the service suite and dynamically react to changes
in the workload and service offerings. 
But exploration under GAs is rather ad-hoc, 
like random search. 

Simulated annealing  (or just annealing)
\cite{AK89,HS89} 
has been widely used for complex,
non-convex optimization problems, including for practical applications,
since its development in the 1980s. Annealing is suitable for
the foregoing resource management problems which operate in a highly dynamic
environment over an indefinite time horizon. Annealing
can more quickly react to detected changes in operating conditions by 
simply increasing its temperature parameter to more aggressively (broadly) search the parameters it controls
(and/or by expanding its ``local neighborhood" sets).
Annealing's local neighborhood set can be adjusted to 
accommodate rules of thumb. 
Also, annealing can be hybridized with a
GA or random search to improve breadth of search,
or hybridized with, e.g., a Tabu search mechanism to improve local-search efficiency.
Thus, we herein employ annealing as a representative method of online optimization.



In this paper, we take an annealing approach  
to the problem of
provisioning an IaaS  cluster and sizing the containers
within its VMs for a plurality of
different streaming workloads.
Annealing works to minimize a macroscopic
objective that can account for factors like current job execution times
and the cost per unit time of the existing cluster.
While it can operate both offline 
(e.g., \cite{TCM21})
and online (runtime), our focus is on the latter.
In addition to HiBench workload case-studies,
we apply an online annealing method
to select the number and sizes of Virtual Machines (VMs)
for a deep-learning workload.
We also apply online annealing to 
the container-sizing problem of a microservice workload
relying on the default container assignment mechanism of
the K8s cluster manager.
Experimental results are reported for these prototypes.

%% file: background.tex
\section{Background}\label{sec:back}

\subsection{Annealing}\label{sec:bg-Annealing}
Simulated annealing was introduced in the 1980s
as a generic
framework to minimize a complicated function $Y:D\rightarrow\mathbb{R}$
over 
a very large discrete bounded domain $D$,
where
``complicated" here means that $Y$ has plural local minima 
in addition
to global ones and $D$ may be a finite discretization of a
continuous domain.

A {\em local}
neighborhood function $\nu(x)$
for all $x\in D$ is defined, where $x\not\in\nu(x)$.
A collection of possible  transitions between $x$ and elements of $\nu(x)$ are also defined,
often taken as all equally likely as assumed in the following (i.e., each with 
uniform probability $1/|\nu(x)|$).
A key requirement of the neighborhood function $\nu$ is that it has
to produce a connected graph in $D$\footnote{That is, the Markov
chain resulting from the ``base" transition probabilities associated with
the neighborhood function is ``irreducible". These transition probabilities should be
chosen so that the base Markov chain is also time-reversible \cite{AK89}.}. Typically,
the neighborhood function ensures only {\em incremental} one-step changes
to the current configuration state, but this is not a requirement.

Given $Y,\nu$, one can define an annealing Markov chain on 
$D$ at temperature $\tau>0$ with 
transition probabilities from
$x$ to $x' \in\nu(x)$ being:
$$ \frac{1}{|\nu(x)|} \exp\left( -\frac{\max\{Y(x')-Y(x),~0\}}{\tau}\right) .$$
We see that a possible transition from the current state $x$
to the next state $x'$ is ``accepted" with positive
probability  even when 
the objective $Y$ is increased, i.e., when $Y(x')>Y(x)$, and
always accepted when $Y(x')\leq Y(x)$ -- this is the
``heat bath" rule.
When the temperature parameter $\tau$ increases,
this acceptance probability increases, i.e.,
there is more exploration and less exploitation which is particularly 
useful when trying to avoid poor local minima.
If the temperature $\tau$ is initially
sufficiently high and slowly
(logarithmically) decreases to zero over time, it can be shown
that the (time-inhomogeneous) Markov chain $Y$ will converge in probability to
its global minimum on $D$ \cite{AK89}.
But this limiting result is not very useful in practice.
Even early on, some authors pointed out that it may be better
not to thus ``cool" the annealing chain \cite{HS89}, particularly
when considering a finite time-horizon. 
If the temperature
is fixed $\tau>0$ and the neighborhoods all have the same size
($|\nu(x)|$ is a constant function of $x\in D$) then (time-homogeneous) 
Markov chain $Y$ has (Gibbs) 
stationary distribution proportional to $\exp(-Y(x)/\tau)$.
Note that as the temperature $\tau\rightarrow 0$, only transitions
that reduce $Y$ are accepted, i.e., pure exploitation. 

In the past, annealing was successfully applied to
complex optimization
problems such as placement and routing 
of VLSI circuits and 
large-scale bin-packing problems \cite{AnnealingResourceAllocation}.

\begin{figure}[ht]
    \centering
    \includegraphics[width=0.45\textwidth]{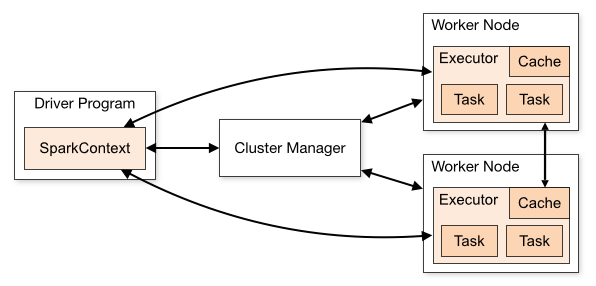}
    \caption{Apache Spark Architecture}
    \label{fig:spark}
\end{figure}
\subsection{Apache Spark}\label{sec:bg-spark}

Apache Spark \cite{spark, spark-the-paper} 
is an open source computing engine used for large-scale data processing applications. A typical Spark cluster setup consist of a single master and one or more workers. When a Spark application is submitted, a Spark driver is launched, which is the responsible component that requests for resources from the cluster manager (e.g., 
YARN, Kubernetes (K8s) or Spark’s Standalone cluster manager) as shown in Figure \ref{fig:spark}. 
Upon arrival of jobs, Spark schedules them in FIFO fashion, where each job is divided into map and reduce stages
(or ``phases"). These stages are represented as a Directed Acyclic Graphs (DAG) i.e., a job's execution plan generated by Spark. Executors are launched on worker nodes and tasks of each stage are sent to run on these executors.
The memory size and number of cores assigned to executors can be modified through a configuration file when an application is submitted. Recent improvements have been made to Spark to better accommodate streaming applications, such as those supported by the alternative platform  Flink \cite{flink}.

\subsection{Kubernetes}

Kubernetes (K8s) is an open-source container orchestration platform. It has gained popularity due to its ability to automate the deployment, scaling, and management of containerized applications in complex distributed systems. A K8s \textit{ deployment} is an abstraction layer that manages the creation, scaling, and updating of \textit{pods}, which are the smallest deployable units in K8s. Deployments provide a declarative way to manage the state of the application, ensuring that the desired number of replicas is always available. Pods, on the other hand, are single instances of a containerized application that run in a shared environment. They provide a lightweight and flexible way to manage containers and their resources. Each pod has a unique IP address and a set of shared resources, including storage volumes and network interfaces. By setting resource requests and limits, K8s can allocate resources more efficiently and ensure that each pod has enough resources to run effectively.

The K8s Application Programming Interface (API) is an interface that enables developers to manage K8s resources by providing a mechanism to create, read, update, and delete resources, such as deployments, pods, and services. To update resources requests and limits, developers can use the K8s API directly or the \texttt{kubectl} command-line tool. The API enables developers to retrieve the current deployment configuration, modify the resources requests and limits in the configuration file, and subsequently update the deployment. The K8s API ensures that the updated deployment is automatically rolled out to the cluster, maintaining the desired state of the system. In this paper, we exploit the API to size the different deployments of microservices with the resource requests of the annealing process.